\documentclass[aps,prl,floatfix,twocolumn,showpacs,reprint,superscriptaddress]{revtex4-1}
\usepackage{amsfonts}
\usepackage{mathrsfs}
\usepackage{amsmath}% needed for subequations
\usepackage{color}
\usepackage{graphicx}
\usepackage{bm}% bold maths
\usepackage{amssymb}
\usepackage{xspace}
\usepackage{epstopdf}
\usepackage{dcolumn}% Align table columns on decimal point
\usepackage{longtable}
\usepackage{multirow}
\usepackage[colorlinks=true, letterpaper=true, pdfstartview=FitV, linkcolor=blue, citecolor=blue, urlcolor=blue]{hyperref}
\begin{document}

\preprint{APS/123-QED}

\title{ Electric Field Influenced Coordinate Jump of the Guiding Center and Magnetotransport}

\author{Jingjing Feng}
\affiliation{Shanghai Institute for Science of Science, Fudan University, Shanghai, China}

\author{Yang Gao}
\affiliation{Department of Physics, The University of Texas at Austin, Austin, Texas 78712, USA}

\author{Qian Niu}
\affiliation{Department of Physics, The University of Texas at Austin, Austin, Texas 78712, USA}

\date{\today}

\begin{abstract}
We derived a electrical current formula in the presence of a strong out-of-plane magnetic field and an in-plane electric field, and within two dimensional disordered system. This current is originated from guiding center coordinate jump 
At strong magnetic field regime $\omega\tau >1$, the current can be pictured as the migration of the center coordinates. During the electron-impurity scattering,  the guiding centers suddenly shift its coordinate. Because of the electric field, the coordinate shift accumulatively contribute to a longitudinal current. 
During the scattering, the value of cyclotron radius changes, which compensates the change of the electric potential energy during the coordinate jump. The diversion of cyclotron radius is the classical manifestation of electric field dependent broadening and shifting of the Landau levels. Our conductivity $\frac{mn}{B^2 \tau}$, derived from direct current response in the linear response regime gives the same result with Kubo's theory derived from fluctuation of current. Our result is valid at one-time collision condition, which gives lower limit of electric field $E>\frac{eB^2 \sqrt{2aR}}{m \pi}$, i.e. $E>500 V/m$ with atomic sized impurity. 
% what are the new point that differentiate the work?? 

\end{abstract}

\pacs{}

\maketitle

{\it Introduction.}--- In condensed matter physics, the problem ``how electrical transport is affected by out-of-plane magnetic field" has been a central topic. On the other hand, the problem ``how mageto-transport is affected by in-plane electric field" arouses less interest but has not been fully understood yet. 
The electric field plays an important role in the magneto-transport by not only providing a driving force for electron, but also breaking symmetry in the system. 
Its various behaviors, such as the electric field tunable band gap in bilayer or multilayer graphene ~\cite{McCann2006, Castro2007, Oostinga2008, Avetisyan2009, Graphene2014}, electric field influenced Landau Level broadening~\cite{Hu1988, Barker1983, Barker1978, Suzuki1989, Arora1987, Balev1986}, electric field shifted ferroelectric phase transition~\cite{Ranjith2006, Bolten2006, Ang2004, Udalov2015}, electric field induced valley polarization~\cite{Islam2018}, etc., contain a wealth of information about the underlying systems. 
However, one of the questions in this field, i.e., how does electric field influence the electron-impurity scattering in magnetic field, has not been fully understood yet. 

%In a system with randomly distributed impurities, if not considering electric field in scattering process, the electron is either localized in a circular orbit without scattering, or localized near one impurity with repeated scattering. While if considering the electric field during the scattering, the electron drifts within a cyclotron orbit, approaches the impurity in the direction perpendicular to both electric and magnetic field, and drifts away after being scattered without localization. The electron will no longer be localized near the same impurity. 

The electron transport in strong magnetic field can be generally described by the migration of guiding center pioneered by Davydov and Pomeranchuk \cite{Davydov1940}. The derivation of coordinate jump of guiding center during scattering is brought up by Kubo \cite{Kubo1959, Kubo1957, Solid State Physics Kubo1969}.  In Kubo's work, the coordinate jump of guiding center has no electric field dependence, because it does not consider electric field during scattering process. When considering no electric field in scattering process, the electron is either localized in a circular orbit without scattering, or localized near one impurity with repeated scattering. In order to avoid localization, Kubo's work implicitly assumes one time scattering condition in each scattering. The magnetoconductivity is calculated by spontaneous fluctuation of guiding center current in the equilibrium state \cite{Kubo1959, Kubo1957, Solid State Physics Kubo1969}. The average current vanishes, but the fluctuation do not. Unlike Kubo's work, we consider the electric field effect during scattering process, therefore, the average current is non zero. The electron drifts on a cyclotron trajectory, approaches the impurity in the direction perpendicular to both electric and magnetic field, and drifts away after being scattered without localization. Our magnetoconductivity is derived by Ohm's law, i.e. the current response to the electric field. Our theory reaches the same conductivity with Kubo's theory \cite{Kubo1959, Kubo1957, Solid State Physics Kubo1969}. Together with Kubo's theory, we provide a proof of the fluctuation-dissipation theorem, i.e. the response of a system to an applied electric field in thermodynamic equilibrium is the same as its response to a spontaneous fluctuation. 

%should mention electric field influenced landau level broadening. 
% whether should mention the relation of our work to the level broadening? 
% what is the difficulty of this work that prevent previous study? 
% what is the conclusion we have? 
%start point, and end point of this paper. start point: how does the electric field influence the magneto-transport. end point: electric field breaks symmetry of radius which corresponds to LL broadening, lead to coordinate jump current. 
% physical deep relation to Drude model

In our work, we derived a formula of electrical current based on Ohm's law in the presence of an external in-plane electric field, a strong out-of-plane magnetic field in two dimensional disordered system. 
During the electron-impurity scattering, the guiding centers suddenly shift its coordinate, which accumulatively contribute to a longitudinal current. 
At the same time, the value of cyclotron radius changes after scattering, which compensates the change of the electric potential energy during the coordinate jump. The diversion of cyclotron radius is the classical manifestation of electric field dependent broadening and shifting of Landau levels. The resulting longitudinal conductivity $\frac{mn}{B^2 \tau}$ has the same form with Drude theory and Kubo's work in the linear response regime of electric field, which is a proof of the fluctuation-dissipation theorem. Our result is valid at strong $B$ field limit and one-time collision condition, which gives lower limit of electric field $E>\frac{eB^2 \sqrt{2aR}}{m \pi}$, i.e. $E>500 V/m$ with atomic sized impurity.
%we derived the electric field dependent coordinate jump of guiding center and its electrical current in classical regime. It yields the magnetoconductivity proportional to the first order coordinate jump. It turns out the same expression with Drude model in strong magnetic field limit. Specifically, the electric field breaks the symmetry of the energy spectrum of electron during the coordinate jump, and leads to the longitudinal current. During the coordinate jump, the cyclotron radius become diverse from single value in order to compensate the energy shift by the electric field. In deeper perspective, the diverse cyclotron radius is the classical manifestation of Landau Level broadening and shifting induced by electric field in quantum picture. 
%The electric field not only broadens the Landau Levels, but also shifts the Landau Levels. 

We show the following results. 1) During the scattering, there is a sudden shift of guiding center coordinates and spherical coordinates $(X,\,Y,\,R,\,\varphi)\rightarrow(X^{\prime},\,Y^{\prime},\,R^{\prime},\,\varphi^{\prime})$. 2) We derive a formula of the longitudinal current density based on Ohm's law and keep the conductivity in linear response regime. 3) The magnetoconductivity has the same form with Drude model in strong magnetic field limit. 4) The cyclotron radius diverse during the scattering due to the potential change by electric field. 5) The energy spectrum after scattering is broadened and shifted due to electric field. This corresponds to the Landau Level asymmetric broadening in quantum picture.

\begin{widetext}
\begin{figure}[b]
\centering
\setlength{\abovecaptionskip}{0pt}
\setlength{\belowcaptionskip}{0pt}
\scalebox{0.21}{\includegraphics*{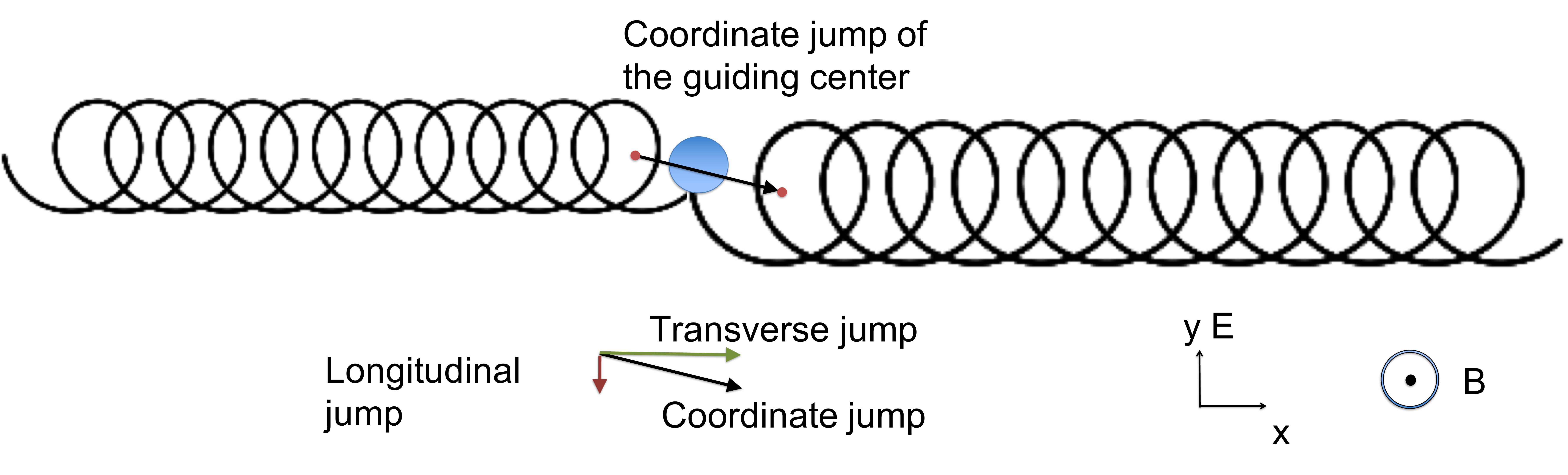}}
\caption{The cyclotron motion of electron and with electron-impurity scattering in two dimension x-y plane. }
\label{fig_sket_imp_scat}
\end{figure}
\end{widetext}

{\it Coordinate Jump of Guiding Center.}--- We start from the classical picture of electron
motion in two dimensional plane with no electron spin and interaction between electrons. The electric
field is in $y$-direction and magnetic field is in $z$-direction perpendicular to the $x-y$ plane, 
where the electrons move, as shown in Fig. \ref{fig_sket_imp_scat}. 
Each of the impurities is randomly and dilutely distributed, therefore, there
is no correlation between impurities. 

The electron motion in electromagnetic field can be described as the superposition of a relatively fast circular motion around guiding center and a relatively slow drift of guiding center. The guiding center drift velocity is $\frac{\mathbf{E} \times \mathbf{B}}{B^{2}}$ in x-direction (perpendicular to both electric and magnetic field), and the relative velocity of electron around the guiding center is $v_{cyc}$(see Fig. \ref{fig_sket_imp_scat}). The guiding center drift velocity is perpendicular to both the electric field and magnetic field. 
Unlike a closed circular orbital velocity, the size of electron velocity in lab frame $v_{lab}$ is changing during the cyclotron motion. $v_{lab}$ is a summation of the velocity of the guiding center $\mathbf{v}_{gc}=\frac{E}{B}\mathbf{x}$ and the relative velocity of electron around the guiding center $\mathbf{v}_{cyc}$:
$\mathbf{v}_{lab}=\mathbf{v}_{cyc}+\mathbf{v}_{gc}$. 

\begin{figure}[h]
\centering
\setlength{\abovecaptionskip}{0pt}
\setlength{\belowcaptionskip}{0pt}
\scalebox{0.4}{\includegraphics*{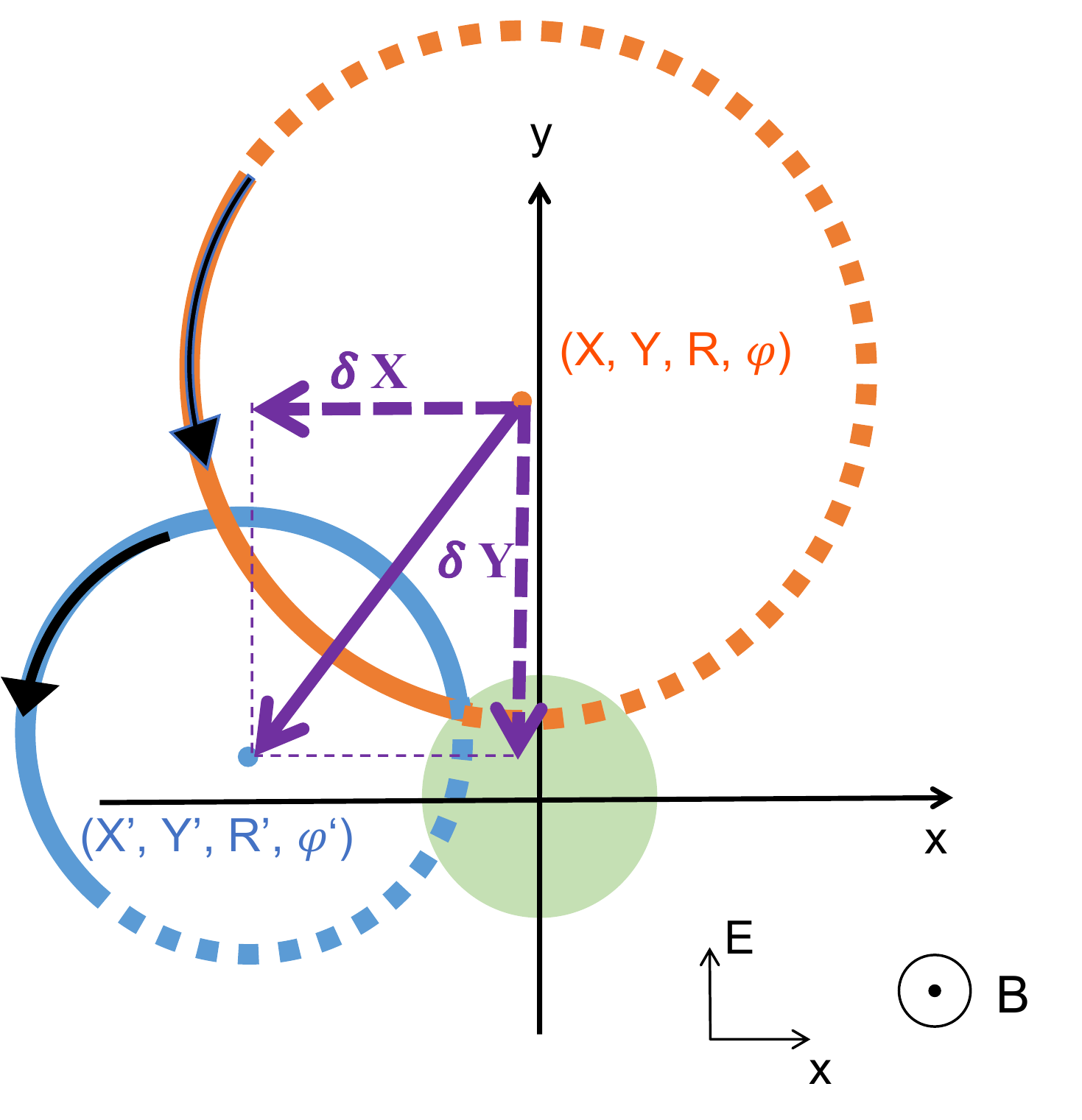}}
\caption{The cyclotron motion of electron and with electron-impurity scattering in two dimension $x-y$ plane. }
\label{fig_R_change}
\end{figure}

In order to study the guiding center motion, we use guiding center coordinate and spherical coordinate $(X\text{,}\,Y,\,R,\,\varphi)$, where $(X\text{,}\,Y)$ are the guiding center coordinates, $R$ is the cyclotron radius, and $\varphi$ is the angle of electron on the circular orbit. Before scattering, the variables $(Y,\,R)$ stays unchanged, and variables $(X,\,\varphi)$ changes with time:  $X=X_0+\frac{E}{B} t$,  $\varphi=\varphi_0+\omega t$, where $X_0$ and $\varphi_0$ are $X(t=0)$ and $\varphi (t=0)$, respectively. During the scattering, there is a sudden shift of all four variables $(X,\,Y,\,R,\,\varphi)\rightarrow(X^{\prime},\,Y^{\prime},\,R^{\prime},\,\varphi^{\prime})$ (see Fig. \ref{fig_R_change}). The shift of the guiding center coordinate $\delta X$, $\delta Y$ is called guiding center coordinate jump (see Fig. \ref{fig_R_change}). 

The shift of cyclotron radius $\delta R$  is due to the energy conservation law at the presence of electric potential energy during scattering (see Fig. \ref{fig_R_change}). With electric field presence at the scattering process, the energy conservation is $ \frac{1}{2}m \omega^2 R^{2} +e\mathbf{E} \cdot \mathbf{Y}=\frac{1}{2}m \omega^2 R^{\prime\,2} +e\mathbf{E} \cdot \mathbf{Y'}$. Because the potential energy shifts along with the guiding center coordinate jump $Y$, the kinetic energy $ \frac{1}{2}m \omega^2 R^{2}$ changes in order to compensate the shift of potential energy $e\mathbf{E} \cdot \mathbf{Y}$. The cyclotron radius is explicitly proportional to the kinetic energy, therefore, the cyclotron radius changes by the scattering.

(In order to clarify the meaning of the electric potential energy, we have the following statement. The guiding center coordinate $Y$ is the average coordinate of electron over one cyclotron period $\overline{Y_{e}}$. The electric potential energy $e\mathbf{E} \cdot \mathbf{Y}$ is thus the average energy of electron over one cyclotron period. )

While in traditional consideration without electric field in scattering process, the energy conservation is $ \frac{1}{2}m \omega^2 R^{2} =\frac{1}{2}m \omega^2 R^{\prime\,2} $, composed of only kinetic energy. Therefore, it requires $R=R'$. 

The expression of guiding center coordinate jump is
\begin{equation}
\delta Y=-R^{\prime} \sin\varphi_{coll}^{\prime}+R\, \sin\varphi_{coll},
\end{equation}
\begin{equation}
\delta X=-R^{\prime} \cos\varphi_{coll}^{\prime}+R\, \cos\varphi_{coll}.
\end{equation}

{\it Longitudinal Current of Guiding Center.}--- 
In this section, we will calculate the steady state current of the system. 
The original condition is that the guiding center, with velocity $\frac{E}{B}$, 
is uniformly distributed in r-space. We now introduce the event line. Due to the steady state condition and the uniformity in real space, the event line abstracts all the scattering events occurring in unit time as taking place on this event line. The event line passes through the center of scatterer and overlaps with $x$ axis. The guiding centers are uniformly distributed on the event line during scattering. When the guiding centers are on the event line, the electrons are uniformly distributed in $Y$ and $\varphi_{0}$ and has $R$ dependence through Fermi distribution. $\varphi_{0}$ is the cyclotron angle when the guiding center reaches the event line, i.e. $\varphi_{0}=\varphi_{coll}+\omega \Delta t$ ($\Delta t$ is the time difference between $\varphi_{0}$ and $\varphi_{coll}$). 
% ? Due to the steady state condition and the uniformity in real space, the event line abstracts all the scattering events occurring in unit time as taking place on this event line. 

In order to calculate the guiding center current, we sum up the guiding center coordinate jump from all the scattering events taking place in unit time. 
 
We start from the derivation of electron distribution function represented by the guiding center coordinate. The density of electrons is 
\begin{equation}
n=\int{f \frac{d^{2}k}{(2\pi)^{2}}}, 
\end{equation} 
where $p$ is the momentum, and $f$ is the Fermi distribution. Using spherical coordinate, 
\begin{equation}
n=\int{ f \frac{k dk d\varphi_0}{(2\pi)^{2}}}. 
\end{equation}

We prove that $\frac{k dk d\varphi_0}{(2\pi)^{2}}$ is equivalent to $ (\frac{eB}{h})^{2} R dR d\varphi_{0}$ represented by guiding center coordinate (Appendix B), therefore, 
\begin{equation}
n=\int{(\frac{eB}{h})^{2} f R dR d\varphi_{0}} \equiv \int{g(R) dR d\varphi_{0}}, 
\end{equation}
where we define the quantity $(\frac{eB}{h})^{2} f R \equiv g(R)$, which is the Fermi distribution function in the guiding center coordinate. 

The number of guiding center crossing the event line per unit time in $dRd\varphi_{0}dY$ is
\begin{equation}
\frac{E}{B}g(R)\,dRd\varphi_{0}dY,
\end{equation} 
where $\frac{E}{B} = \frac{dX}{dt} $, i.e. the guiding center drift velocity along $x$ direction. 

The longitudinal current density along the direction of electric field is
\begin{eqnarray}
j_{y} & = & \sum_{i} {-e n_i v_{i}} \nonumber \\
 & =&  -e \int_{0}^{2\pi}\int_{0}^{R_{F}}g(R) \\
 &  & \times \int_{-(R+a)}^{(R+a)}\,\frac{E}{B} n_{im}\delta Y(R,\varphi_{0},Y)\,dRd\varphi_{0}dY, \nonumber
\label{currentjy1}
\end{eqnarray}
where $n_{im}$ is the impurity density, the subscript $i$ denotes $i$th scattering event. 
% explain the meaning of integrate all deltaY. 

The integral $\frac{E}{B} \int_{-(R+a)}^{(R+a)}\,\int_{0}^{2\pi}\int_{0}^{R_{F}}g(R)\nonumber \times\delta Y(R,\varphi_{0},Y)\,dRd\varphi_{0}dY$ in Eq. \ref{currentjy1} is the summation of all the coordinate jump occurring in unit time at one impurity. The nonzero longitudinal current density $j_{y}$ indicates the average coordinate of the guiding center shifts from $\sum_{i}Y_{i}=0$ to $ \sum_{i}Y^{\prime}_{i} \neq 0$ during scattering. 

In order to solve Eq. \ref{currentjy1}, we firstly derive $\delta Y(R,\varphi_{0},Y)$. However, it is hard to derive $\delta Y$ as a function of $(R,\,\varphi_{0},\,Y)$ analytically. Instead, we derive $\delta Y$ as a function of $(R,\,\varphi_{im},\,\theta_{v})$
\begin{eqnarray}
&  & \delta Y(R,\varphi_{im},\theta_{v})  = \nonumber \\
 &  &  \frac{1}{B\omega_{c}} [ E \cos\theta_{v}+\sqrt{(-E^{2}+B^{2} v_{cyc}^{2}+E^{2} \cos^{2}\theta_{v})} ] \\
 &  & \times ( -2 \cos(\theta_{v}-\varphi_{im} ) \cos\varphi_{im} ), 
\label{deltaY}
\end{eqnarray}
where $\varphi_{im}$
is the angle (starting from $x$-axis) on the impurity when the scattering takes place; the $\theta_{v}$ is the
angle of the incident velocity of electron (starting from $x$-axis). The derivation of $\delta Y(R,\varphi_{im},\theta_{v})$ is in the Appendix \ref{APP-A}. 

However, the distribution function of electron is not uniform along $\varphi_{im}$ and $\theta_{v}$. Therefore, 
we first calculate the weighting factor of $\varphi_{im}$ and $\theta_{v}$, respectively. The pair of variables $(\varphi_{0},\,Y)$ can be transformed to pair of variables $(\varphi_{im},\,\theta_{v})$ by Jacobian determinant $\left[ \begin{array}{cc}
\frac{\partial\varphi_{0}}{\partial\varphi_{im}} & \frac{\partial\varphi_{0}}{\partial\theta_{v}}\\
\frac{\partial Y}{\partial\varphi_{im}} & \frac{\partial Y}{\partial\theta_{v}}
\end{array} \right]$, which determines the weighting factor. 

The Jacobian determinant can be calculated by the following three
equations 
\begin{equation}
Y_{0}=a \sin\varphi_{im}-R \sin\varphi_{coll}, 
\end{equation}
\begin{equation}
\varphi_{0}=\varphi_{coll}-\frac{a \cos\varphi_{im}-R \cos\varphi_{coll}}{\Delta X},
\end{equation}
\begin{equation}
tan\theta_{v}=\frac{v_{cyc} \cos\varphi_{coll}}{-v_{cyc} \sin\varphi_{coll}+\frac{E}{B}},  
\end{equation}
where $\Delta X=\frac{E}{B\omega_{c}}$, which is the drift distance of guiding center after one cyclotron period.  
The range of $\theta_{v}$ in the integral is restricted by the range $[\varphi_{im}+\frac{\pi}{2},\,\varphi_{im}+\frac{3\pi}{2}]$. 

Integrating over $dRd\varphi_{im}d\theta_{v}$, the longitudinal current density in Eq. \ref{currentjy1}
becomes
\begin{eqnarray}
j_{y}  &=&  -en_{im}\frac{E}{B}\int_{0}^{2\pi}\int_{\varphi_{im}+\frac{\pi}{2}}^{\varphi_{im}+\frac{3\pi}{2}}\int_{0}^{R_{F}}\,g(R)\,\delta Y(R,\varphi_{im},\theta_{v})\nonumber \\
 &  & \times \left[ \begin{array}{cc}
\frac{\partial\varphi_{0}}{\partial\varphi_{im}} & \frac{\partial\varphi_{0}}{\partial\theta_{v}}\\
\frac{\partial Y}{\partial\varphi_{im}} & \frac{\partial Y}{\partial\theta_{v}}
\end{array}\right] \,dRd\theta_{v}d\varphi_{im}.
\label{currentjy2}
\end{eqnarray}

By solving Eq. \ref{currentjy2}, the longitudinal current density to the first order of electric field is
\begin{equation}
j_{y}=en_{im}\frac{E}{B}(\frac{eB}{h})^{2}(8a\pi)\frac{1}{3}(\frac{\hbar k_{f}}{eB})^{3}. 
\end{equation}
Note that $\delta Y$ in our theory has electric field dependence (as seen in Eq. \ref{deltaY}) and can be expanded with respect to electric field
\begin{eqnarray}
 &  & \delta Y= -\frac{\sqrt{B^2 {R_0}^2 {\omega_c}^2} (\cos{\theta_v}+\cos{(\theta_v-2 \varphi_{im})})}{B \omega_c} \\
 &  & - \frac{\cos{\theta_v} (\cos{\theta_v}+\cos{(\theta_v-2 \varphi_{im})})}{B \omega_c} E+O\left(  E^{2}\right) 
\end{eqnarray}
Because the Jacobian determinant has $E^{-1}$ and $E^{0}$ terms, we keep only $E^{0}$ and $E^{1}$ terms in $\delta Y$ in order to keep the current $j_y$ in linear regime. (Note, the $E^{-1}$ term in Jacobian determinant times $E^{0}$ term in $\delta Y$ produces $E^{-1}$ term, which however, will vanish after integration in Eq. \ref{currentjy2}. )

The conductivity thus
is $\sigma_{yy}=\frac{8\pi}{3}\frac{e^{2}}{h}(n_{im}a\lambda_{f})(\frac{nh}{eB})^{2}$,
which can be transformed to
\begin{equation}
\sigma_{yy}= \frac{mn}{B^2 \tau},
\end{equation}
based on the transport relaxation time $\frac{1}{\tau}=\frac{8}{3} n_{i} v a$ at $B=0$, and the electron density $n=\frac{k_{f}^2}{4 \pi}$. 

%\begin{equation}
%\sigma_{yy}=\frac{8\pi}{3}\frac{e^{2}}{h}\frac{\left(\frac{A_{q}}{\lambda_{f}^{2}}\right){}^{2}}{\frac{l}{\lambda_{f}}}\,4\pi^{2},
%\end{equation}
%where $A_{q}=\frac{h}{eB}$, is the area between each of the Landau
%levels; $l=\tau v=\frac{8}{3} \tau v$, is the mean free path at $B=0$, where $\tau$ is the transport relaxation time. The longitudinal conductivity can be written as 

{\it Transverse Current of Guiding Center.}--- 
The transverse current density along the direction of electric field is 
\begin{eqnarray}
j_{x} & = & \sum_{i} {-e n_i v_{i} } \nonumber \\
 & =&  -en_{im}\frac{E}{B}\int_{-(R+a)}^{(R+a)}\,\int_{0}^{2\pi}\int_{0}^{R_{F}}g(R)\nonumber \\
 &  & \times\delta X(R,\varphi_{0},Y)\,dRd\varphi_{0}dY. 
\label{currentjx1}
\end{eqnarray}

Same as how we derive $\delta Y(R,\,\varphi_{im},\,\theta_{v})$, we derive $\delta X(R,\,\varphi_{im},\,\theta_{v})$
\begin{eqnarray}
&  & \delta X(R,\varphi_{im},\theta_{v})  = \nonumber \\
 &  &  \frac{1}{B\omega_{c}} [ E \cos\theta_{v}+\sqrt{(-E^{2}+B^{2} v_{cyc}^{2}+E^{2} \cos^{2}\theta_{v})} ] \\
 &  & \times 2 \cos(\theta_{v}-\varphi_{im} ) \sin\varphi_{im} , 
\label{deltaX}
\end{eqnarray}
The derivation of $\delta X(R,\varphi_{im},\theta_{v})$ is in Appendix \ref{APP-A}. 

Integrating over $dRd\varphi_{im}d\theta_{v}$, the transverse current density in Eq. \ref{currentjx1}
becomes
\begin{eqnarray}
j_{x}  &=&  -en_{im}\frac{E}{B}\int_{0}^{2\pi}\int_{\varphi_{im}+\frac{\pi}{2}}^{\varphi_{im}+\frac{3\pi}{2}}\int_{0}^{R_{F}}\,g(R)\,\delta X(R,\varphi_{im},\theta_{v})\nonumber \\
 &  & \times \left[ \begin{array}{cc}
\frac{\partial\varphi_{0}}{\partial\varphi_{im}} & \frac{\partial\varphi_{0}}{\partial\theta_{v}}\\
\frac{\partial Y}{\partial\varphi_{im}} & \frac{\partial Y}{\partial\theta_{v}}
\end{array}\right] \,dRd\theta_{v}d\varphi_{im}.
\label{currentjx2}
\end{eqnarray}

By solving Eq. \ref{currentjx2}, the transverse current density is 
\begin{equation}
j_{x}=0. 
\end{equation}
There is no anomalous component in transverse current. The only transverse current is from drift current of guiding center.

{\it Discussion--Proof of fluctuation-dissipation theorem}--- We provide a simple demonstration of fluctuation-dissipation theorem below. The quantum counterpart of our theory (shown below), combining with Kubo's current fluctuation theory \cite{Kubo1959, Kubo1957, Solid State Physics Kubo1969} is a good demonstration of the fluctuation-dissipation theorem. (The electric field is along $x$ direction in Kubo's theory. For consistency, we keep  the electric field along $y$ direction and the notation of longitudinal coordinate jump as $Y'-Y$ in our manuscript.)

In quantum counterpart of our theory, the longitudinal current is expressed as 
\begin{eqnarray}
j_y=-e \sum_{N,N'} \sum_{Y,Y'} 2 f_N W_{Nk,N'k'} (Y'-Y) ,
\label{jquantum0}
\end{eqnarray}
where $f_N$ is the Fermi distribution function, $W_{Nk,N'k'}$ is scattering probability, i.e. $W_{NY,N'Y'}=\frac  {2\pi }{\hbar } |\langle \psi_{N',Y'}|U|\psi_{N,Y}\rangle|^2 \delta (\varepsilon_{{N',Y'}}-\varepsilon_{N,Y})$, and the factor $2$ is due to spin degeneracy considered in our quantum theory, as well as Kubo's theory. 

By exchanging $N,Y \rightarrow N',Y'$, 
\begin{eqnarray}
j_y=-e \sum_{N',N} \sum_{Y',Y} 2 f_N' W_{N'k',Nk} (Y-Y'),
\label{jquantum00}
\end{eqnarray}
the result of the equation remains the same because $W_{NY,N'Y'}=W_{N'Y',NY}$. Therefore, by adding Eq. \ref{jquantum0} and Eq. \ref{jquantum00} up, the current becomes
\begin{eqnarray}
j_y=-\frac{e}{2} \sum_{N',N} \sum_{Y',Y} 2 (f_N-f_N') W_{N'k',Nk} (Y'-Y),
\label{jquantum000}
\end{eqnarray}

There are three quantities modified by electric field, eigenfunction in $W_{Nk,N'k'}$, eigenenergy $\varepsilon_{N,Y}$ and the energy level density $\delta (\varepsilon_{{N',Y'}}-\varepsilon_{N,Y})$. We will show that Kubo's current fluctuation corresponds to our linear response current originated from the electric field shifting of Landau level. 

The Hamiltonian of an electron under an out of plane magnetic field ($z$ direction) and an in plane electric field ($y$ direction) in disordered system is 
\begin{equation}
H=\frac{(\mathbf{P}+e\mathbf{A})^{2}}{2m}-e\mathbf{E}\cdot \mathbf{y}+U. 
\end{equation}

The Schordinger equation of electron in this system is 
\begin{equation}
\{\frac{(\mathbf{P}+e\mathbf{A})^{2}}{2m}-e\mathbf{E}\cdot \mathbf{y}+U\}\psi(x, y)=2m\varepsilon \psi(x, y). 
\end{equation}
Assuming that the solution is $\psi(x, y)=e^{ik_x x} \psi(y)$, and choosing Landau Gauge $A_x=-B y$, $A_y=A_z=0$, substitute this solution to the Schordinger equation, we get
\begin{equation}
\{\frac{1}{2m} (P_x-eBy)^{2}-eEy+U\}\psi(y)=2m\varepsilon \psi(y). 
\end{equation}

The above equation can be transformed as 
\begin{eqnarray}
 \{-\frac{\partial^{2}}{\partial y^{2}}&+&e^{2}B^{2}[y-(\frac{\hbar k_{x}}{eB}+\frac{mE}{eB^{2}})]^{2} \\ \nonumber
&-&2m[\frac{mE^{2}}{2B^{2}}+\frac{E \hbar k_{x}}{B}] +2mU\}\psi(y)=2m\varepsilon \psi(y). 
\label{schordinger}
\end{eqnarray}
It has been defined that $Y \equiv  l^2 k_{x}$ (where $l$ is the magnetic length, i.e. $l=\sqrt{\frac{\hbar}{m\omega}}$), which is the guiding center of cyclotron without electric field, and is a good quantum number. In our case, the position of guiding center is shifted by the inclusion of electric field, i.e. $Y \rightarrow (Y+\frac{mE}{eB^{2}})$. The eigenstate is also modified by electric field as $\psi_N(Y+\frac{mE}{eB^{2}})$. The eigenenergy becomes $\varepsilon_{N,Y}=\hbar\omega(N+\frac{1}{2})-eE Y \equiv \varepsilon_N-eE Y$. The distribution function $f_N= \frac {1}{e^{[\hbar \omega (N+\frac{1}{2})-\mu ]/k_{B}T}+1}$ remains the same because both the eigenenergy and the chemical potential is shifted by $eE Y$, which cancels out. 

Therefore, Eq. \ref{jquantum000} becomes 
\begin{eqnarray}
j_y=&& -\frac{e}{2} \sum_{N',N} \sum_{Y',Y} 2 (f(\varepsilon_N)-f(\varepsilon_{N'})) \nonumber \\
&& \cdot \frac{2\pi }{\hbar } |\langle \psi_{N',Y'}|U|\psi_{N,Y}\rangle|^2 \delta [\varepsilon_{N'}-\varepsilon_{N}- eE(Y'-Y)] \nonumber \\
&& \cdot(Y-Y'),  
\end{eqnarray}
because of energy conservation, $\varepsilon_{N'}=\varepsilon_{N}+ eE(Y'-Y)$, $f(\varepsilon_{N'}) \equiv f(\varepsilon_{N}+ eE(Y'-Y))$. 

As long as $E$ is small so that $eE(Y'-Y)/k_B T \ll \hbar \omega$, the distribution function $f(\varepsilon_{N}+ eE(Y'-Y))$ can be expanded at $E= 0$, we get
\begin{eqnarray}
f(\varepsilon_{N}+&& eE(Y'-Y))= f(\varepsilon_{N}) \\
&& +\frac{\partial f(\varepsilon_{N})}{\partial \varepsilon_{N}} eE(Y'-Y)+O\left(  E^{2}\right).   
\end{eqnarray}
Therefore, the longitudinal current becomes 
\begin{eqnarray}
j_y=&& \frac{e}{2} \sum_{N',N} \sum_{Y',Y} 2 \frac{\partial f(\varepsilon_{N})}{\partial \varepsilon_{N}} eE(Y'-Y)^2 \nonumber \\
&& \cdot \frac{2\pi }{\hbar } |\langle \psi_{N',Y'}|U|\psi_{N,Y}\rangle|^2 \delta [\varepsilon_{N'}-\varepsilon_{N}- eE(Y'-Y)] .   
\end{eqnarray}

The current density is $j_y/V$, where $V$ is the volume of the material. 

This linear response current corresponds to the conductivity derived from fluctuation of current in Kubo' theory. 

In Kubo's theory, the conductivity is 
 \begin{eqnarray}
 \sigma_{yy}= &  & \frac{2e^2}{V} \sum_{N, Y, p_z} \sum_{N', Y', p_z'} {\frac{\partial f(\varepsilon_N (p_z))}{\partial \varepsilon_N (p_z)} } \\\nonumber
 & & \cdot \frac{1}{2} (Y-Y')^2 W_{N'Y'p_z',NYp_z}. 
\label{Kubo_conductivity}
\end{eqnarray}
Because the system is 2D in our case, by separating $p_z$, we get
 \begin{eqnarray}
 &  & \sigma_{yy}= \frac{2e^2}{V} \sum_{N, Y} \sum_{N', Y'} {\frac{\partial f(\varepsilon_N)}{\partial \varepsilon_N} }\frac{1}{2} (Y-Y')^2 W_{N'Y',NY}. 
\label{Kubo_conductivity}
\end{eqnarray}

%For short range scattering potential, the longitudinal conductivity by solving Eq. \ref{Kubo_conductivity} is
% \begin{eqnarray}
%\sigma_{xx}= \frac{n e^2}{kT} \frac{1}{2} l^2 n_s 4\pi f^2 \hbar \Omega \langle 1/|p_z|\rangle,   
%\end{eqnarray}
%where $\langle 1/|p_z|\rangle$ is an average of the reverse of momentum in $z$ direction $1/|p_z|$ of incident electrons. 
%As $\frac{1}{\tau}=n_s 4\pi f^2 \hbar \Omega ( 1/|p_z|)$, the longitudinal conductivity becomes
%\begin{eqnarray}
%\sigma_{xx}=\frac{mn}{B^2 \tau}. 
%\end{eqnarray}
%
%Our longitudinal conductivity is derived by averaging the current and is derived from Ohm's law $j_y=env_y$. Our longitudinal conductivity is proportional to the first order of coordinate jump. 

Both of our theory and Kubo's theory gives conductivity $\frac{mn}{B^2 \tau}$. Combining Our theory and Kubo's theory, it is an specific example to prove the fluctuation-dissipation theorem, i.e. the response of a system to an applied electric field in thermodynamic equilibrium is the same as its response to a spontaneous fluctuation.

{\it Discussion--Comparison with Drude Theory}--- Traditional Drude theory considers the electron-impurity scattering as a friction force macroscopically. The equation of motion is%
\begin{equation}
m\mathbf{\dot{v}}=-e(\mathbf{E}+\mathbf{v\times \mathbf{B}})-\frac{m \mathbf{v}}{\tau},
\end{equation}
where $v$ is the average velocity per electron, $\tau$ is the mean time an electron has traveled since the last collision. 

It yields the relationship between longitudinal current density $\mathbf{J}$ and electric field $\mathbf{E}$, 
\begin{equation}
\mathbf{J}=\frac{\frac{ne^{2}\tau}{m}}{1+(\frac{eB}{m})^{2}\tau^{2}} \mathbf{E}. 
\end{equation}

Under strong magnetic field limit $\omega \tau \gg 1$, the longitudinal current density becomes 
\begin{equation}
\mathbf{J}=\frac{mn}{B^2 \tau} \mathbf{E}.
\label{ohm}
\end{equation}

In our theory, we look into each scattering process in detail and provide a microscopic method to calculate the current. We first brought up the strong magnetic field limit. Based on this limit, the electron motion can be represented by guiding center motion. Then, we figured out that each scattering process can be pictured as a guiding center coordinate jump. By accumulating all the coordinate jumps, we derived the current formula Eq. \ref{currentjy1} which gives the strong field conductivity. Because of the consideration of detailed scattering process, we provide an explicit expression for reverse of relaxation time $\frac{1}{\tau}=\frac{8}{3} n_{i} v a$ and electron density $n=\frac{k_{f}^2}{4 \pi}$. 

%The relation between the longitudinal current density and electric field in Eq. \ref{ohm} is a macroscopic explanation of Ohm's law $\mathbf{J}=-en \mathbf{v}$. The electron kinetic energy in strong magnetic field can be considered as the kinetic energy of guiding center $\frac{1}{2}m\frac{E^2}{B^2}$. The electron potential energy averaged over the times of electrons they had since their last collision is $-eEv \left\langle t \right\rangle$, where $t$ is the time since that collision. At any particular moment in time, $\left\langle t \right\rangle$ is half of the mean time between collisions $\tau$: $\left\langle t \right\rangle = \frac{\tau}{2}$. Because the energy is conserved, $-eEv \left\langle t \right\rangle= \frac{1}{2}m\frac{E^2}{B^2}$. Thus, $v= \frac{m }{e B^2 \tau} E$. Therefore, $\mathbf{J}=-en \mathbf{v}=-\frac{m n}{ B^2 \tau} \mathbf{E}$. 
% ?don't understand why tau/2? 

%Our theory is an microscopic way to reach the result of Drude theory in strong magnetic field limit. Our theory also applies Ohm's law in Eq. \ref{currentjy1}. However, we provide a microscopic method to calculate the electron velocity based on the coordinate jump velocity during the collision. 
% key differentiates our theory from Drude theory? 
% how could the result be the same? not directly visible. the similarity behind the not straight forward coincidence? 

% physical relation to Drude model. 
% mention hall current=0
% what is the cj velocity here? how to derive? 
% what is the scattering probability here? 

{\it Discussion--Change of Cyclotron Radius During Scattering.}--- 
The cyclotron radius is changed after each collision (illustrated in Fig. \ref{fig_sket_imp_scat}). Macroscopically, it is due to electric field potential change during the coordinate jump, as we mentioned in section 'Coordinate Jump of Guiding Center'. 

Microscopically, at the moment of collision, the incident velocity of electron is the sum of guiding center velocity and the relative velocity of electron $\mathbf{v}_{in}=\mathbf{v}_{gc}+\mathbf{v}_{cyc}=(-v_{cyc} \sin\varphi_{coll}+\frac{E}{B},\,v_{cyc} \cos\varphi_{coll})$, where $\varphi_{coll}$ is the  angle on the cyclotron orbit (starting from x-axis) at the incident moment. 

At the moment after collision, the outgoing velocity of electron is
$\mathbf{v}_{out}=\mathbf{v}_{gc}+\mathbf{v}_{cyc}^{\prime}=(-v_{cyc}^{\prime} \sin\varphi_{coll}^{\prime}+\frac{E}{B},\,v_{cyc}^{\prime} \cos\varphi_{coll}^{\prime})$, where $\varphi_{coll}^{\prime}$ is the angle on the cyclotron orbit  (starting from x-axis) at the moment after collision. 

Because of energy conservation, the value of velocity at the moment of collision is a constant, i.e. $|\mathbf{v}_{in}|=|\mathbf{v}_{out}|$. However, the cyclotron velocity before and after collision are different because
\begin{eqnarray}
v_{cyc}^{\prime\,2}&=&(v_{out,x}-\frac{E}{B})^2+v_{out,y}^2\\
&=& v_{out,x}^2+v_{out,y}^2+\frac{E^2}{B^2}-2 v_{out,x} \frac{E}{B}\\
&=& v_{in,x}^2+v_{in,y}^2+\frac{E^2}{B^2}-2 v_{out,x} \frac{E}{B}\\
&=& (-v_{cyc} \sin{\varphi}+\frac{E}{B})^2+v_{cyc}^2 \cos{\varphi}^2 + \frac{E^2}{B^2}\\
&& -2 v_{out,x}\frac{E}{B}\\
&=& v_{cyc}^2+2 \frac{E^2}{B^2} - 2v_{cyc} \frac{E}{B} \sin{\varphi} -2 v_{out,x}\frac{E}{B}. 
\end{eqnarray}
 
Therefore, the difference between the  velocity square before and after scattering is
\begin{equation}
v_{cyc}^{\prime\,2}-v_{cyc}^{2}=2 \frac{E^2}{B^2} - 2v_{cyc} \frac{E}{B} \sin{\varphi} -2 v_{out,x}\frac{E}{B}. 
\end{equation}
Thus, the change of kinetic energy is
\begin{equation}
\triangle\varepsilon_{kin}=\frac{1}{2}m(v_{cyc}^{\prime\,2}-v_{cyc}^{2}). 
\end{equation}

 {\it Discussion--Asymmetric Landau Level Broadening.}--- 
 As we mentioned below Eq. \ref{currentjy1}, the total coordinate jump in unit time is $\frac{E}{B} \int_{-(R+a)}^{(R+a)}\,\int_{0}^{2\pi}\int_{0}^{R_{F}}g(R)\nonumber \times\delta Y(R,\varphi_{0},Y)\,dRd\varphi_{0}dY = \frac{E}{B}(\frac{eB}{h})^{2}(8a\pi)\frac{1}{3}(\frac{\hbar k_{f}}{eB})^{3}$. The potential energy changes by 
 \begin{equation}
\sum_{i} eE \cdot \delta Y_{i} = eE \frac{E}{B}(\frac{eB}{h})^{2}(8a\pi)\frac{1}{3}(\frac{\hbar k_{f}}{eB})^{3}. 
\end{equation}
The same goes with the kinetic energy 
\begin{equation}
\sum_{i} \Delta \varepsilon_{kin}=- eE \frac{E}{B}(\frac{eB}{h})^{2}(8a\pi)\frac{1}{3}(\frac{\hbar k_{f}}{eB})^{3}. 
 \end{equation}
The kinetic energy spectrum of electron after scattering is not only broadened, but also asymmetrically broadened with respect to the original kinetic energy before scattering, which leads to a shift in its average energy level.

%From another perspective, the energy conservation of guiding center is equivalent with the energy conservation of electron in the co-moving frame of the guiding center.  The co-moving frame of the guiding center moves at the same velocity with the guiding center drift velocity. At the co-moving frame, the effective electric field $E=0$, the electron trajectory is a closed circular orbit due to the Lorentz force, with the velocity of electron $\mathbf{v}_{cyc}$. The impurities are moving in the opposite direction of the co-moving frame. The kinetic energy of electron is $\frac{1}{2}m v_{cyc}^{2}$. The 

%This is the conservation law that relates the motion of the guiding
%center and the electron. However it is not obvious how the energy
%of the electron and the energy of the guiding center are related.
%The guiding center is the position of an electron averaged over one
%period, and the jump of the guiding center $\delta Y$ can be seen
%as the jump of electrons averaged over one period $\overline{\delta Y_{e}}$,
%that is $\delta Y=\overline{\delta Y_{e}}$. The above equation (9)
%is the conservation law of electron motion averaged in one period. 

{\it Condition for One-Time Collision.}--- Our theory is valid when the electron only collides once on a impurity
before colliding with another impurity. The impurity has the size
of an atom, which is $a\sim10^{-9}m$; the cyclotron radius is approximately
$10^{-6}m$ in 2DEG, relatively large compared with the impurity.
Because of this, it is reasonable to consider each of the electrons
only collides once on one impurity and then scattered away. The $\triangle X$,
that is $\triangle X=\frac{E}{B\omega_{c}}$, the distance that the
guiding center moves after one cycle of the cyclotron motion, has
to be large enough, $\varDelta X\gg a$, in order to have an electron
collide only once on one impurity everytime before colliding with
another impurity. This generates a lower limit to the range of electric
field. 

We quantitatively derive the condition for one-time-collision as $\triangle X\geq10^{-8}m$,
or say $E\geq500\,V/m$ if $B\sim0.5\,Tesla$. We use the following
way. The area of the guiding centers, in which all possilbe collisions
will take place, is a circlewise ring within radius R-a and R+a. After
the first collision, the guiding center is inside the ring, and will
keep moving in x direction, it will or will not again pass through
the area of the ring depending on where the first collision takes
place. The sufficient and necessary condition of no second collision
is the electron will not be at the certain angle on cyclotron motion
which superposes the impurity during its second pass through the area
of the ring. We divide this area of ring into four parts: A, B, C
and D (fig. 3). Only the condition for area B and C needs to be considered,
as they give a sufficient lower limit which satisfies the condition
for A and D. 
% plot one-time collision plot! 

\vskip 0.8cm
\begin{acknowledgments}
We acknowledge useful discussions with Qi Chen, Liang Du, Liuyang Sun, Cong Xiao, Zhi Wang, Chao Lei, Ming Xie, Haodi Liu, Kaige Hu, Ming Sun. Q.N. is supported by DOE (DE-FG03-02ER45958, Division of Materials Science and Engineering) in the formulation of our theoy. J.F. and Y.G. are supported by NSF (EFMA- 1641101) and Welch Foundation (F-1255). 
\end{acknowledgments}

\vskip0.8cm
\appendix
\label{appendix}

\section{\\Appendix A: \\
Derivation of coordinate jump $\delta Y(R,\varphi_{im},\theta_{v})$}
\label{APP-A}

As discussed in previous section, the incident velocity of electron $\mathbf{v}_{in}$ and the outgoing velocity $\mathbf{v}_{out}$ have the same absolute value. 

On the other hand, the direction of outgoing velocity is to turn the incident velocity counter-clockwise by angle $\theta$, where $\theta=\theta_{v_{out}}- \theta_{v_{in}}=2\varphi_{im}-2\theta_v-\pi$. Because the incident velocity of electron is $\mathbf{v}_{in}=(-v_{cyc} \sin\varphi_{coll}+\frac{E}{B},\,v_{cyc} \cos\varphi_{coll})$, the outgoing velocity can be expressed as
\begin{eqnarray}
\mathbf{v}_{out} &=& \left[ \begin{array}{cc}
\cos{\theta} & -\sin{\theta} \\
\sin{\theta} & \cos{\theta}
\end{array}\right] 
\left[ \begin{array}{cc}
-v_{cyc} \sin\varphi_{coll}+\frac{E}{B} \\
v_{cyc} \cos\varphi_{coll}
\end{array}\right]  \\
&=& \left[ \begin{array}{cc}
(-v_{cyc} \sin\varphi_{coll}+\frac{E}{B})\cos{\theta}- v_{cyc} \cos\varphi_{coll}\sin{\theta}\\
(-v_{cyc} \sin\varphi_{coll}+\frac{E}{B})\sin{\theta}+ v_{cyc} \cos\varphi_{coll}\cos{\theta}
\end{array}\right]. 
\end{eqnarray}

As discussed in previous section, the outgoing velocity expressed in terms of $v_{cyc}^{\prime}$ and $\varphi_{coll}^{\prime}$ is $\mathbf{v}_{out}=(-v_{cyc}^{\prime} \sin\varphi_{coll}^{\prime}+\frac{E}{B},\,v_{cyc}^{\prime} \cos\varphi_{coll}^{\prime})$. Combining the two expressions of $\mathbf{v}_{out}$, the relationship between $\varphi_{coll}^{\prime}$ and $\varphi_{coll}$, $\theta$ is 
\begin{eqnarray}
-v_{cyc}^{\prime} \sin\varphi_{coll}^{\prime}+\frac{E}{B}&=&(-v_{cyc} \sin\varphi_{coll}+\frac{E}{B})\cos{\theta}\\
&&- v_{cyc} \cos\varphi_{coll}\sin{\theta}, 
\label{vpsin}
\end{eqnarray}
and
\begin{eqnarray}
v_{cyc}^{\prime} \cos\varphi_{coll}^{\prime}&=&(-v_{cyc} \sin\varphi_{coll}+\frac{E}{B})\sin{\theta}\\
&&+ v_{cyc} \cos\varphi_{coll}\cos{\theta}.  
\label{vpcos}
\end{eqnarray}

As we know $\delta Y = -R^{\prime} \sin\varphi_{coll}^{\prime}+R\, \sin\varphi_{coll}$, our goal is to express $R^{\prime}$, $\varphi_{coll}$ and $\varphi_{coll}^{\prime}$ in terms of $(\varphi_{im},\theta_{v})$, in order to express $\delta Y$ by $(\varphi_{im},\theta_{v})$. 

From $\mathbf{v}_{in}=(-v_{cyc} \sin\varphi_{coll}+\frac{E}{B},\,v_{cyc} \cos\varphi_{coll})$, there is
\begin{eqnarray}
\cos \theta_v=\frac{-v_{cyc} \sin\varphi_{coll}+\frac{E}{B}}{v_{in}}, 
\end{eqnarray}
and
\begin{eqnarray}
\sin \theta_v=\frac{v_{cyc} \cos\varphi_{coll}}{v_{in}},   
\end{eqnarray}
where $v_{in}=\sqrt{v_{cyc}^2+\frac{E^2}{B^2}-2 v_{cyc} \frac{E}{B} \sin\varphi_{coll}}$. To solve for $\sin\varphi_{coll}$ and $\cos\varphi_{coll}$, there is 
\begin{eqnarray}
&v_{cyc} \sin\varphi_{coll}&=\frac{E}{B}-\frac{E}{B} \cos^2 \theta_v\\
&&-\frac{1}{B} \cos \theta_v \sqrt{-E^2+E^2 \cos^2 \theta_v +B^2 v_{cyc}^2}, 
\label{vsin}
\end{eqnarray}
and
\begin{eqnarray}
v_{cyc} \cos\varphi_{coll}&=&\sin \theta_v [ \sqrt{v_{cyc}^2+\frac{E^2}{B^2}\cos^2 \theta_v-\frac{E^2}{B^2}}\\
& & +\sqrt{\frac{E^2}{B^2}\cos^2 \theta_v} ].  
\label{vcos}
\end{eqnarray}

Combining Eq. {vpsin}, Eq. {vpcos}, Eq. {vsin} and Eq. {vcos}, we can reach the goal to express $R^{\prime}$, $\varphi_{coll}$ and $\varphi_{coll}^{\prime}$ in terms of $(\varphi_{im},\theta_{v})$. 

Therefore, we finally express $\delta Y $ in terms of $(\varphi_{im},\theta_{v})$
\begin{eqnarray}
\delta Y(R,\varphi_{im},\theta_{v})  &  &= -\frac{1}{B\omega_{c}^{2}} [ E\omega_{c} \cos\theta_{v}\nonumber \\
 &  & +\sqrt{\omega_{c}^{2}(-E^{2}+B^{2}v_{cyc}^{2}+E^{2} \cos^{2}\theta_{v})} ] \\
 &  & \times2 \cos\theta_{v}\cos^{2}(\theta_{v}-\varphi_{im})\nonumber \\
 &  & -\frac{1}{B\omega_{c}^{2}} [ \sqrt{(E\omega_{c}\cos\theta_{v})^{2}}\nonumber \\
 &  & +\sqrt{\omega_{c}^{2}(-E^{2}+B^{2}v_{cyc}^{2}+E^{2}\cos^{2}\theta_{v})} ] \nonumber \\
 &  & \times \sin\theta_{v} \sin(2\theta_{v}-2\varphi_{im}), 
\end{eqnarray}

\section{\\Appendix B: \\
Connection between real space integral and the momentum space integral}
\label{APP-B}

We will prove that the momentum space integral $\frac{k dk d\varphi_0}{(2\pi)^{2}}$ is equivalent to the real space integral $ (\frac{eB}{h})^{2} R dR d\varphi_{0}$, where $\varphi_0$ is the angle of momentum electron on the cyclotron orbit with respect to $x$ axis. 

First, $dx dy=J_1 R dR d\varphi_0$, where 
\begin{eqnarray}
\displaystyle J_1={\begin{vmatrix}{\frac {\partial x}{\partial R}}&{\frac {\partial x}{\partial \varphi_0 }}\\[2pt]{\frac {\partial y}{\partial R}}&{\frac {\partial y}{\partial \varphi_0 }}\end{vmatrix}}={\begin{vmatrix}\cos \varphi_0 &-R\sin \varphi_0 \\\sin \varphi_0 &R\cos \varphi_0 \end{vmatrix}}=R. 
\end{eqnarray}
because $x=R \cos \varphi_0$, and $y=R \sin \varphi_0$. Therefore, $dx dy=R dR d\varphi_0$. 

Second, it can be proven that $\frac{dk_x dk_y}{(2\pi)^{2}}=J_2 \frac{dk d\varphi_0}{(2\pi)^{2}}$, where 
\begin{eqnarray}
\displaystyle J_2={\begin{vmatrix}{\frac {\partial k_x}{\partial k}}&{\frac {\partial k_x}{\partial \varphi_0 }}\\[2pt]{\frac {\partial k_y}{\partial k}}&{\frac {\partial k_y}{\partial \varphi_0 }}\end{vmatrix}}={\begin{vmatrix}-\sin \varphi_0 & -k\cos \varphi_0 \\\cos \varphi_0 & -k\sin \varphi_0 \end{vmatrix}}=k, 
\end{eqnarray}
because $k_x=-k \sin \varphi_0$, and $k_y=-k \cos \varphi_0$. Therefore, $\frac{dk_x dk_y}{(2\pi)^{2}}=k \frac{dk d\varphi_0}{(2\pi)^{2}}$. 

In addition, because $k=\frac{mv}{\hbar}=\frac{eB R}{\hbar}$, $k \frac{dk d\varphi_0}{(2\pi)^{2}}= (\frac{eB}{h})^{2} R dR d\varphi_{0}$.

\bibliographystyle{apsrev4-1}

\end{document}